\documentclass[12pt]{article}
\newcommand{\bq}{\begin{equation}}
\newcommand{\eq}{\end{equation}}
\newcommand{\ba}{\begin{eqnarray}}
\newcommand{\ea}{\end{eqnarray}}

\newcommand{\nl }{ \nonumber  }

\newcommand{\p}{\partial}
\newcommand{\pu}{\p_\tau}

\newcommand{\h}{\hspace{1cm}}
\newcommand{\s}{\sigma}

\begin{document}
\begin{flushright}
{\bf hep-th/9911210}
\end{flushright}
\vspace*{1.cm}
{\bf\begin{center}
  NULL BRANES IN STRING THEORY BACKGROUNDS
\vspace*{0.5cm}\\ P. Bozhilov \footnote {E-mail:
bojilov@thsun1.jinr.ru}, \\ \it Bogoliubov Laboratory of Theoretical
Physics, \\ JINR, 141980 Dubna, Russia
\end{center}}
We consider null bosonic $p$-branes moving in curved space-times and
develop a method for solving their equations of motion and
constraints, which is suitable for string theory backgrounds. As an
application, we give an exact solution for such background in ten
dimensions.
\\

PACS number(s): 11.10.Lm, 04.25.-g, 11.27.+d



\section{Introduction}
\hspace{1cm} The null $p$-branes are the zero tension limit $T_p\to
0$ of the usual $p$-branes, the 1-brane being a string. This
relationship between them generalizes the correspondence between
massless and massive particles. Thus, the tensionless branes may be
viewed as a high energy limit of the tensile ones.

The $p$-branes are characterized by an energy scale $T_p^{1/(p+1)}$
and therefore by a length scale $T_p^{-1/(p+1)}$. The gravitational
field provides another length scale, the curvature radius of the
space-time $R_c$. For a $p$-brane moving in a gravitational field an
appropriate parameter is the dimensionless constant ${\cal D}=R_c
T_p^{1/(p+1)}$. Large values of ${\cal D}$ imply weak gravitational
field. One may reach such values of ${\cal D}$ by letting $T_p \to
{\infty}$. In the opposite limit, small values of ${\cal D}$, one
encounters {\it strong gravitational fields} and it is appropriate to
consider $T_p \to 0$, i.e. {\it null} or {\it tensionless} branes.

A Lagrangian which could describe under certain conditions null
bosonic branes in $D$-dimensional Minkowski space-time was first
proposed in \cite{BLS862}. An action for a tensionless $p$-brane with
space-time supersymmetry was first given in \cite{Z8812}. Since then,
other types of actions and Hamiltonians (with and without
supersymmetry) have been introduced and studied in the literature
\cite{G92,BZ93,ILST93,HLU94,PS97,B986,B991,B992}. Owing to their zero
tension, the worldvolume of the null $p$-branes is a light-like,
$(p+1)$-dimensional hypersurface, imbedded in the Minkowski
space-time. Correspondingly, the determinant of the induced metric is
zero. As in the tensile case, the null brane actions can be written
in reparametrization and space-time conformally invariant form.
However, their distinguishing feature is that at the classical level
they may have any number of global space-time supersymmetries and be
$\kappa$-invariant in all dimensions, which support Majorana (or
Weyl) spinors. At the quantum level, they are anomaly free and do not
exhibit any critical dimension, when appropriately chosen operator
ordering is applied \cite{BZ89,G92,BZ93,PS97,B9711}. The only
exception are the tensionless branes with manifest conformal
invariance, with critical dimension $D=2$ for the bosonic case and
$D=2-2N$ for the spinning case, $N$ being the number of worldvolume
supersymmetries \cite{PS97}.

Let us mention also the paper \cite{GRRA892}, which is devoted to the
construction of field theory propagators of null strings and
$p$-branes, as well as the corresponding spinning versions.

Almost all of the above investigations deal with {\it free} null
branes moving in {\it flat} background (a qualitative consideration
of null $p$-brane interacting with a scalar field has been done in
\cite{BZ93}). The interaction of tensionless membranes ($p=2$) with
antisymmetric background tensor field in four dimensional Minkowski
space, described by means of Wess-Zumino-like action, is studied in
\cite{Z8990}. The resulting equations of motion are successfully
integrated exactly.

To our knowledge, the only papers till now devoted to the classical
dynamics of {\it null} $p$-branes ($p>1$) moving in {\it curved}
space-times are \cite{RZ96,B993,BD99}.

In \cite{RZ96}, the null $p$-branes living in $D$-dimensional
Friedmann-Robertson-Walker space-time with flat space-like section
(${\it k}=0$) have been investigated. The corresponding equations of
motion have been solved exactly. It was argued that an ideal fluid of
null $p$-branes may be considered as a source of gravity for
Friedmann-Robertson-Walker universes.

In \cite{B993}, the classical mechanics of null branes in a
gravity background was formulated. The Batalin-Fradkin-Vilkovisky
approach in its Hamiltonian version was applied to the considered
dynamical system. Some exact solutions of the equations of motion and
of the constraints for the null membrane ($p=2$) in general stationary
axially symmetric four dimensional gravity background were found. The
examples of Minkowski, (A)dS, Schwarzschild, Taub-NUT and Kerr
space-times were considered. Another exact solution, for the
Demianski-Newman background, can be found in \cite{BD99}.

In this article we consider the classical evolution of tensionless
bosonic $p$-branes in a particular type of $D$-dimensional curved background.
In Sec. 2 we develop a method for solving their equations of motion and
constraints. In Sec. 3, as an application of the method proposed,
we give an explicit exact solution for the ten dimensional solitonic
five-brane gravity background. Sec. 4 is devoted to our concluding remarks.

\section{Solving the equations of motion}
\hspace{1cm} We will use the following reparametrization invariant action
for the null bosonic $p$-brane living in a $D$-dimensional curved space-time
with metric tensor $g_{MN}(x)$:
\ba\label{a} S=\int d^{p+1}\xi {\cal L} \h,\h
{\cal L}=V^mV^n\p_m x^M\p_n x^N g_{MN}(x),\h\h
\\ \nl
\p_m=\p/\p\xi^m \h, \h \xi^m=(\xi^0,\xi^a)=(\tau,\s^a),\h\h
\\ \nl
m,n=0,1,...,p \h,\h a,b=1,...,p \h,\h M,N=0,1,...,D-1. \ea It is a
natural generalization of the flat space-time action given in
\cite{ILST93,HLU94}.

Let us rewrite the Lagrangian density from (\ref{a}) into the form
($\pu=\p/\p\tau, \p_a=\p/\p\s^a$): \ba\label{L} {\cal
L}=\frac{1}{4\lambda^0} g_{MN}(x)\bigl (\pu-\lambda^a\p_a\bigr )x^M
\bigl (\pu-\lambda^b\p_b\bigr )x^N , \ea where the connection between
$V^m$ and $(\lambda^0,\lambda^a)$ is given by \ba\nl
V^m=\bigl(V^0,V^a\bigr)=\Biggl(-\frac{1}{2\sqrt{\lambda^0}},
\frac{\lambda^a}{2\sqrt{\lambda^0}}\Biggr) . \ea The
Euler-Lagrange equation for $x^K$ are \ba\label{eqx}
\pu\Biggl [\frac{1}{2\lambda^0}\bigl (\pu-\lambda^b\p_b\bigr
)x^{K}\Biggr ] -\p_a\Biggl [\frac{\lambda^{a}}{2\lambda^{0}}\bigl
(\pu-\lambda^b\p_b\bigr )x^{K} \Biggr ] \\ \nl +
\frac{1}{2\lambda^0}\Gamma^{K}_{MN} \bigl (\pu-\lambda^a\p_a\bigr
)x^M \bigl (\pu-\lambda^b\p_b\bigr )x^N = 0 , \ea
where $\Gamma^{K}_{MN}$ is the connection compatible with the metric
$g_{MN}(x)$: \ba\nl \Gamma^{K}_{MN}=\frac{1}{2}g^{KL}\bigl(
\p_{M}g_{NL}+\p_{N}g_{ML}-\p_{L}g_{MN}\bigr) . \ea
The equations of motion for the Lagrange multipliers $\lambda^{0}$
and $\lambda^{a}$ which follow from (\ref{L}) give the constraints:
\ba\label{Tx1}
g_{MN}(x)\bigl (\pu-\lambda^a\p_a\bigr )x^M \bigl
(\pu-\lambda^b\p_b\bigr )x^N = 0 , \\ \label{Tx2} g_{MN}(x)\bigl
(\pu-\lambda^b\p_b\bigr )x^M \p_a x^N = 0 . \ea

From now on, we will work in the gauge $\lambda^0,
\lambda^a = constants$, in which the equations
(\ref{eqx}) have the form : \ba\label{eqxf}
\bigl(\pu-\lambda^a\p_a\bigr)\bigl (\pu-\lambda^b\p_b\bigr )x^{K} +
\Gamma^{K}_{MN} \bigl (\pu-\lambda^a\p_a\bigr )x^M \bigl
(\pu-\lambda^b\p_b\bigr )x^N = 0 . \ea We are going to look for
solutions of the equations of motion (\ref{eqxf}) and constraints
(\ref{Tx1}), (\ref{Tx2}) for the following type of gravity background
\ba\label{gm} ds^2 &=&g_{MN}dx^M dx^N\\ \nl &=& g_{qq}(dx^q)^2 +
2g_{qs}dx^q dx^s + g_{ss}(dx^s)^2 + g_{\alpha\beta}dx^\alpha dx^\beta
+ \sum_{i}g_{ii}(dx^i)^2 ,\ea
where $x^q\neq x^s$ are two arbitrary
coordinates and it is supposed that $g_{MN}$ does not depend on them.

To simplify the equations of motion (\ref{eqxf}) and constraints
(\ref{Tx1}), (\ref{Tx2}), we introduce the ansatz \ba\label{az} x^q
(\tau,\sigma^a)&=&C^q F(z^a) + y^q (\tau),\h  x^s
(\tau,\sigma^a)=C^s F(z^a) + y^s (\tau),\\ \nl x^M
(\tau,\s^a)&=&y^M (\tau)\h\mbox{for}\h M\neq q,s;
\h z^a =\lambda^{a}\tau+\s^a ,\ea where $F(z^a)$ is an
arbitrary function of $z^a$, and $C^q$, $C^s$ are constants.
Inserting (\ref{az}) in (\ref{Tx1}),
(\ref{Tx2}) and (\ref{eqxf}), one obtains (the dot is used for
$d/d\tau$) \ba\label{ey} \ddot y^K + \Gamma^{K}_{MN}\dot y^M \dot y^N
= 0,\\ \label{c1} g_{MN}\dot y^M \dot y^N = 0,\\ \label{c2} \bigl(C^q
g_{qq} + C^s g_{qs}\bigr)\dot y^q + \bigl(C^q g_{qs} + C^s
g_{ss}\bigr)\dot y^s = 0.\ea It turns out that for the given metric
(\ref{gm}), the equations for $\dot y^q$ and $\dot y^s$ in (\ref{ey})
become linear differential equations. On the other hand, with the
help of (\ref{c2}), we can separate the variables in them. The
corresponding solution, compatible with (\ref{c2}), is
$(C=const)$\ba\nl \dot y^q(\tau) = -C\left(C^q g_{qs}+C^s
g_{ss}\right)\exp(-{\cal H}) ,
\\ \label{sqs}
\dot y^s(\tau)=+C\left(C^q g_{qq}+C^s g_{qs}\right)\exp(-{\cal H}) ,
\\ \nl
{\cal H} = \int\Bigl(g^{qq}dg_{qq}+2g^{qs}dg_{qs}+g^{ss}dg_{ss}\Bigr)
. \ea Now we observe that if we introduce the matrix \ba\nl {\bf h}
=
\left(\begin{array}{cc}g_{qq}&g_{qs}\\g_{qs}&g_{ss}\end{array}\right)
,\ea then the following equality holds \ba\label{h} \exp\left({\cal
H}\right) = \exp\left(\int Tr{\bf h}^{-1}d{\bf h}\right) =
\mbox{det}{\bf h}\equiv h .\ea At the same time, the above equality
is the compatibility condition for the solution (\ref{sqs}) with the
other equations of motion and constraint (\ref{c1}).

Using (\ref{sqs}) and (\ref{h}), the equations for the other
coordinates and the remaining constraint can be rewritten as
($K,M,N\neq q,s$) \ba\nl \ddot y^K + \Gamma^{K}_{MN}\dot y^M \dot y^N
+ \frac{1}{2}g^{KM}\p_M \left(C^2 \frac{G}{h}\right) = 0,\\
\label{c1a} g_{MN}\dot y^M \dot y^N + \left(C^2 \frac{G}{h}\right) =
0,\ea where \ba\nl G = (C^q)^2 g_{qq}+2C^q C^s g_{qs}+(C^s)^2
g_{ss}.\ea

At this stage, taking into account the general structure of the
string theory gravity backgrounds in $D$ dimensions, we introduce an
additional restriction on the metric (\ref{gm}). Namely, we suppose
that the set of values of the index $M$ is expressed by the subsets
$M=(q,s,\alpha,i)$ such that $g_{MN}$ does not depend on coordinates
$x^{\alpha}(\tau,\s^a) = y^{\alpha}(\tau)$ in addition to
$x^q$, $x^s$. Under this condition, one can reduce the order of the
differential equations for $y^{\alpha}$ by one with the result
\ba\label{alpha} \dot y^{\alpha} = \exp\left(- \int
g^{\alpha\beta}dg_{\beta\gamma}\right)C^{\gamma}, \h C^{\gamma}=constants.\ea
The condition on (\ref{alpha}) to be in accordance with the equations for
$y^{i}$ and the constraint (\ref{c1a}) is \ba\nl
2\frac{dg_{\alpha\beta}}{d\tau}\dot y^{\alpha}\dot y^{\beta} +
g_{\alpha\beta}\frac{d}{d\tau}\left(\dot y^{\alpha}\dot
y^{\beta}\right) = 0,\ea and it is identically satisfied.

Let us turn to the equations of motion for the remaining coordinates
$y^{i}$. These are: \ba\label{eyi} \ddot y^i + \Gamma^{i}_{jk}\dot
y^j \dot y^k + \frac{1}{2}g^{ij}\left[\p_j \left(C^2
\frac{G}{h}\right) - \left(\p_j g_{\alpha\beta}\right) \dot
y^{\alpha}\dot y^{\beta}\right] = 0,\ea where $\dot{y}^{\alpha}$ are given
by (\ref{alpha}). The following step is to use
the equality \ba\nl \left(\p_j g_{\alpha\beta}\right) \dot
y^{\alpha}\dot y^{\beta} = - \p_j\left(g_{\alpha\beta}\dot
y^{\alpha}\dot y^{\beta}\right),\ea which is an identity on the
solutions (\ref{alpha}). This allow us to transform the equations
(\ref{eyi}) into the form: \ba\nl
2\frac{d}{d\tau}\left(g_{ij}\dot{y}^{j}\right) - \left(\p_i
g_{jk}\right)\dot y^j \dot y^k + \p_i\left(C^2\frac{G}{h} +
V\right) = 0,\ea where
\ba\nl V = \left(C\exp\left(\int gdg^{-1}\right)\right)^{\alpha}
g_{\alpha\beta}\left(\exp\left(-\int g^{-1}dg\right)C\right)^{\beta}.\ea
Taking into account that the matrix $g_{ij}$ is a diagonal one, we can
further transform the equations for $y^i(\tau)$ to obtain (there is
no summation over $i$):
\ba\label{eyif}\frac{d}{d\tau}\left(g_{ii}\dot{y}^{i}\right)^2+
\dot{y}^{i}\p_i\left[g_{ii}\left(C^2\frac{G}{h} +V\right)\right]+
\dot{y}^{i}\sum_{j\ne i}\p_i \left(\frac{g_{ii}}{g_{jj}}\right)
\left(g_{jj}\dot{y}^{j}\right)^2 = 0.\ea In receiving (\ref{eyif}),
the constraint (\ref{c1a}) rewritten in the form
\ba\label{c1f} g_{ii}(\dot{y}^{i})^2 + \sum_{j\ne
i}g_{jj}(\dot{y}^{j})^2 + V + C^2\frac{G}{h} = 0 \ea is also used.

Now it is evident from (\ref{eyif}) that we can reduce the order of
these differential equations by one, if \ba\label{tf} \p_i
\left(\frac{g_{ii}}{g_{jj}}\right) = 0 \h\mbox{for}\h i\ne j\ea or
\ba\label{yd}\p_i \left(g_{jj}\dot{y}^{j}\right)^2 = 0 \h\mbox{for}\h
i\ne j . \ea Keeping in mind the aim to apply our results to the
string theory gravity backgrounds, we choose the following
combination of the two existing possibilities: for all coordinates
$y^i$ except one, which we call $y^r$, the equalities (\ref{tf}) are
fulfilled; for $i=r$, the equalities (\ref{yd}) hold. Then the
result of integration, compatible with (\ref{c1f}), is \ba\label{fik}
\left(g_{kk}\dot{y}^k\right)^2 &=& C_k
\left(y^r,...,y^{k-1},y^{k+1},...\right) - g_{kk}\left(C^2\frac{G}{h}
+ V \right)\\ \nl &=& E_k \left(...,y^{k-1},y^k,y^{k+1},...\right), \\
\label{fid} \left(g_{rr}\dot{y}^r\right)^2 &=&
g_{rr}\left\{\left(\sum_{k}-1\right) \left(C^2\frac{G}{h} + V \right)
- \sum_{k}\frac{C_k}{g_{kk}} \right\} \\ \nl &=& E_r (y^r),
\\ \label{cds} \p_k \left(\frac{g_{kk}}{g_{ii}}\right) &=& 0,
\h k\neq q,s,\alpha,r.\ea Here $C_k$, $E_k$ and $E_r$ are arbitrary
functions. $C_k$ depend on all coordinates on which depends the
metric, but $y^k$ (for every fixed value of $k$). $E_k$ do not depend
on $y^r$, but depend on $y^k$. Obviously, the right hand sides of
(\ref{fik}) and (\ref{fid}) have to be nonnegative.

Now, we are interested in finding exact solutions of the above
equations. It turns out that it is preferable to use a slightly
different approach for multi dimensional and for lower dimensional
space-times. This will allow us to obtain solutions in more general
class of metrics in the lower dimensional case. At first, we will try
to find solutions appropriate for application to higher dimensional
backgrounds.

A simple analysis shows that we can integrate the equations
(\ref{fik}) and (\ref{fid}) completely, if we fix the
coordinates on which the background depends, except $y^r$. We
prefer to consider just this possibility in connection with further
applications in mind. Because $g_{MN}=g_{MN}(y^r,y^k)$,
($k\ne q,s,\alpha,r$), we fix the coordinates $y^k$: $y^k =
y^{k}_{0} =constants$. Then the {\it exact} solution of the equations
of motion and constraints for a null $p$-brane in this background is
given by (\ref{az}), where $y^k$ are constants and \ba\nl y^q = y^q_0
\mp C \int_{y^r_0}^{y^r}du\frac{\left(C^q g^0_{qs}+C^s
g^0_{ss}\right)}{h^0 W^{1/2}_0},\h y^q_0,y^r_0=const;\\ \nl y^s = y^s_0
\pm C \int_{y^r_0}^{y^r}du \frac{\left(C^q g^0_{qq}+C^s
g^0_{qs}\right)}{h^0 W^{1/2}_0},\h y^s_0=const;
\\ \label{exs} y^{\alpha} = y^{\alpha}_0
\pm C^{\gamma}\int_{y^r_0}^{y^r}\frac{du}{W^{1/2}_0}
\exp\left(-\int dg^0_{\gamma\beta}g^{0\beta\alpha}\right)
,\h y^{\alpha}_0=const;\\ \nl \tau  = \tau_{0} \pm
\int_{y^r_0}^{y^r}\frac{du}{W^{1/2}_0} ,\h \tau_{0}=const;
\\ \nl W_0 = -\frac{C^0_k}{g_{rr}^0 g^0_{kk}} ,
\h C^0_{k} = g^0_{kk}
\left(C^2\frac{G^0}{h^0} + V^0 \right).\ea In
the above equalities $g^0_{MN} = g^0_{MN}(y^r) = g_{MN}(y^r, y^k_0)$
and analogously for $G^0$, $h^0$, $V^0$ and $C^0_{k}$.

Let us turn to the lower dimensional case. This separate
consideration is necessary, because in obtaining the solution
(\ref{exs}) we have restricted the metric to be independent on too
many variables. In four dimensions for instance, it is preferable to
have a metric, which depends at least on two of the coordinates. This
gives us the possibility to consider different types of black hole
backgrounds for example. Taking this into account, now we would like
to find exact solutions of the differential equations (\ref{fik}),
(\ref{fid}) for background metric, which does not depend only on
$y^q$ and $y^s$. To this end, we set $\alpha =\{\emptyset\}$ and
choose the coordinates $y^k$, $(k\ne q,s,r)$ to be constant. Then for
the remaining coordinates one obtains \ba\nl y^q = y^q_0 \mp
\int_{y^r_0}^{y^r}du \left(C^q g^0_{qs}+C^s
g^0_{ss}\right)\left[-\frac{g^0_{rr}}{G^0 h^0}\right]^{1/2}, \\
\label{exsl} y^s = y^s_0 \pm\int_{y^r_0}^{y^r}du \left(C^q
g^0_{qq}+C^s g^0_{qs}\right)\left[-\frac{g^0_{rr}}{G^0
h^0}\right]^{1/2} \\ \nl \tau  = \tau_{0} \pm \int_{y^r_0}^{y^r}du
\left[-\frac{g^0_{rr}h^0}{C^2 G^0} \right]^{1/2}.\ea
The corresponding {\it exact} solution for the tensionless $p$-brane
in the chosen $D$ - dimensional gravity background is given again by
(\ref{az}) with (\ref{exsl}) inserted in there.

Finally, we note that for obtaining exact solutions in cosmological
type backgrounds, one can identify $x^0$ with $y^r$ in (\ref{exs}) or
in (\ref{exsl}).

\section{The explicit solution}
\hspace{1cm}In this section we are going to apply the method proposed
in the previous one for finding an explicit exact solution. We start
by considering a null $p$-brane moving in the solitonic
$(\tilde{d}-1)$-brane background \cite{DKL95} \ba\nl ds^2 =
g_{MN}dx^M dx^N = \exp\left(2A\right)\eta_{\mu\nu}dx^{\mu}dx^{\nu} +
\exp\left(2B\right)\left(dr^2 + r^2d\Omega^2_{D-\tilde{d}-1}\right),
\\ \nl \exp\left(2A\right) =
\left(1+\frac{k_{\tilde{d}}}{r^d}\right)^{-\frac{d}{d+\tilde{d}}}, \h
\exp\left(2B\right) =
\left(1+\frac{k_{\tilde{d}}}{r^d}\right)^{+\frac{\tilde{d}}{d+\tilde{d}}},
\h k_{\tilde{d}}=const, \\ \nl d + \tilde{d} = D-2,\h
\eta_{\mu\nu}=\mbox{diag}(-,+,...,+), \h \mu, \nu =
0,1,...,\tilde{d}-1. \ea The $(D-\tilde{d}-1)$-dimensional sphere
$\mathbf{S}^{D-\tilde{d}-1}$ is supposed to be parameterized so that
\ba\nl g_{kk} &=&
\exp\left(2B\right)r^2\prod_{n=1}^{D-k-1}\sin^2\theta_{n},\h D-k-1 =
1,2,...,D-\tilde{d}-2, \\ \nl g_{D-1,D-1} &=&
\exp\left(2B\right)r^2.\ea If we now set $q=0$, $\alpha =
1,2,...,\tilde{d}-2, D-1$, $s=\tilde{d}-1$, $r=\tilde{d}$, $k=\tilde{d} +
1,...,D-2$, the conditions (\ref{cds}) are fulfilled and we can use
the general formula (\ref{exs}). The result is
\ba\nl y^{\mu} = y^{\mu}_0 \pm E^{\mu}
\int_{r_0}^{r}du\left(1+\frac{k_{\tilde{d}}}{u^d}\right)\left(
\mathcal{E}-\frac{\left(C^{D-1}\right)^2}{u^2}+\frac{\mathcal{E}k_{\tilde{d}}}{u^d}
\right)^{-1/2},\h r\equiv y^{\tilde{d}},\\ \nl E^{\mu}=
\left(E^0,E^1,\ldots,E^{\tilde{d}-1}\right)=
\left(CC^{\tilde{d}-1},C^1,\ldots,CC^0\right); \\ \label{exsdv}
\varphi = \varphi_0 \pm C^{D-1}
\int_{r_0}^{r}\frac{du}{u^2}\left( \mathcal{E}-
\frac{\left(C^{D-1}\right)^2}{u^2}
+ \frac{\mathcal{E}k_{\tilde{d}}}{u^d} \right)^{-1/2},\h
\varphi\equiv y^{D-1};\\ \nl \tau = \tau_0 \pm
\int_{r_0}^{r}du\left(1+\frac{k_{\tilde{d}}}{u^d}\right)^
{\frac{\tilde{d}}{d+\tilde{d}}} \left(
\mathcal{E}-\frac{\left(C^{D-1}\right)^2}{u^2} +
\frac{\mathcal{E}k_{\tilde{d}}}{u^d} \right)^{-1/2};\h\h\hspace{.5cm}
\\ \nl \mathcal{E}\equiv - E^{\mu}E^{\nu}\eta_{\mu\nu}
= \left(CC^{\tilde{d}-1}\right)^2 -
\left(CC^0\right)^2 - \sum_{\alpha =
1}^{\tilde{d}-2}\left(C^{\alpha}\right)^2\geq 0.\ea

Let us restrict ourselves to the particular case of ten dimensional
solitonic 5-brane background. The corresponding values of the
parameters $D$, $\tilde{d}$ and $d$ are $D=10$, $\tilde{d}=6$, $d=2$.
Taking this into account and performing the integration in
(\ref{exsdv}), one obtains the following explicit exact solution of
the equations of motion and constraints for a tensionless $p$-brane
living in such curved space-time \ba\nl x^0 (\tau,\s^a)&=&C^0
F(z^a) + y^0 (\tau),\\ \nl  x^5 (\tau,\s^a)&=&C^5
F(z^a) + y^5 (\tau),\\ \nl x^M (\tau,\s^a)&=&y^M
(\tau) \h\mbox{for}\h M\neq 0,5,7,8,\\ \nl x^{7,8}
(\tau,\s^a)&=&y^{7,8}_0 = constants,\ea where for
$\mathcal{C} = k_6 - \left(C^{9}\right)^2/\mathcal{E} > 0$ \ba\nl y^{\mu} =
y^{\mu}_0 \mp \frac{k_6 E^{\mu}}{\left(\mathcal{CE}\right)^{1/2}}
\ln\left(\frac{\frac{\mathcal{C}^{1/2}}{r} + \left(1 +
\frac{\mathcal{C}}{r^2}\right)^{1/2}}{\frac{\mathcal{C}^{1/2}}{r_0} +
\left(1 + \frac{\mathcal{C}}{r_{0}^{2}}\right)^{1/2}}\right) \\ \nl
\pm \frac{E^{\mu}}{\mathcal{E}^{1/2}}\left[\left(\mathcal{C} +
r^2\right)^{1/2} - \left(\mathcal{C} + r_{0}^{2}\right)^{1/2}\right],
\\ \nl \varphi = \varphi_0 \mp
\frac{C^9}{\left(\mathcal{CE}\right)^{1/2}}
\ln\left(\frac{\frac{\mathcal{C}^{1/2}}{r} + \left(1 +
\frac{\mathcal{C}}{r^2}\right)^{1/2}}{\frac{\mathcal{C}^{1/2}}{r_0} +
\left(1 + \frac{\mathcal{C}}{r_{0}^{2}}\right)^{1/2}}\right), \\
\label{expls} \tau = \tau_0 \mp \left(\frac{k_6^3}{\mathcal{C}^4
\mathcal{E}^2}\right)^{1/4}r^{3/2}\left(1 +
\frac{\mathcal{C}}{r^2}\right)^{1/2}F_2\left(3/4,1,-3/4;3/2,3/4;1 +
\frac{r^2}{\mathcal{C}},-\frac{r^2}{k_6}\right)\\ \nl \mp
2\left(\frac{k_6^3 r_0^2}{\mathcal{C}^2 \mathcal{E}^2}\right)^{1/4}
F_1\left(1/4,1/2,-3/4;5/4;-\frac{r_0^2}{\mathcal{C}},
-\frac{r_0^2}{k_6}\right)\\ \nl \pm
\frac{\Gamma(1/4)k_6}{4\Gamma(3/4)}\sqrt{\frac{\pi}{\mathcal{CE}}}
\mbox{\scriptsize{2}}F_{1}\left(1/4,1/2;-1/4;1 -
\frac{k_6}{\mathcal{C}}\right)\\ \nl \mp
\frac{\Gamma(1/4)k_6}{2\Gamma(3/4)}\sqrt{\frac{\pi}{\mathcal{CE}}}
\left(1 - \frac{k_6}{\mathcal{C}}
\right)^{-1/4}\mbox{\scriptsize{2}}F_{1}\left(1/4,3/2;3/4;\left(1 -
\frac{k_6}{\mathcal{C}}\right)^{-1}\right). \ea In the above
expressions, $\Gamma(z)$ is the Euler's $\Gamma$-function and
$\mbox{\scriptsize{2}}F_{1}\left(a,b;c;z\right)$ is the Gauss'
hypergeometric function. The functions $F_1(a,b,b';c;w,z)$ and
$F_2(a,b,b';c,c';w,z)$ in (\ref{expls}) are two of the hypergeometric
functions of two variables. The defining equalities for $F_1$ and $F_2$
are \cite{GR,PBM} \ba\nl F_1(a,b,b';c;w,z) =
\sum_{k,l=0}^{\infty}\frac{(a)_{k+l} (b)_{k}
(b')_{l}}{k!l!(c)_{k+l}}w^k z^l ,\h (|w|,|z|<1);\\ \nl
F_2(a,b,b';c,c';w,z) = \sum_{k,l=0}^{\infty}\frac{(a)_{k+l} (b)_{k}
(b')_{l}}{k!l!(c)_{k}(c')_{l}}w^k z^l ,\h (|w|+|z|<1), \ea where
\ba\nl (a)_{k} = \frac{\Gamma (a+k)}{\Gamma (a)}.\ea

\section{Concluding remarks}
\hspace{1cm} In this paper we performed some investigation on the
classical dynamics of the null bosonic branes in a curved space-time.
In the second section, we have found {\it exact} solutions
of the equations of motion and constraints for a
null $p$-brane in two particular types of $D$-dimensional curved
backgrounds. However, the latter are general enough to include in
itself many interesting cases of string theory gravity backgrounds in
different dimensions (like black branes, intersecting branes and
cosmological type backgrounds). In the third section, we gave an explicit
example of exact solution for the solitonic 5-brane curved background
in ten dimensions.

Let us briefly comment on the area of applicability of the obtained results.
Considering this, we have to take into account the following properties
of the tensionless extended objects:
\begin{enumerate}
\item
the null $p$-branes can be considered as a high energy limit of the
tensile ones, when the role played by the string tension may be ignored;
\item
in the presence of strong gravitational fields, it is appropriate to
consider the null tension limit of a brane;
\item
the large tension limit of a $q$-brane is related to the zero
tension limit of the dual $p$-brane.
\end{enumerate}
For example, if the null branes are viewed as space-time probes,
the obtained exact solutions may have relevance to the singularity
structure of branes. On the other hand, these solutions may
have cosmological implications especially in the early universe.
It is worth checking if this type of solutions leads to self-consistent
brane cosmology. Another appropriate field of application of our results
is the investigation of the solution properties near black hole horizons.

Outside the framework of the exact solutions, one can try to find an
approximate solution for a tensile $p$-brane by perturbative expansion
in powers of the brane tension.
Then the exact null brane solution will be the zero approximation.
However, it is more interesting to answer the question: can we calculate all
the terms in such an expansion? In other words: does our method work in the
tensile brane case? It turns out that the answer is positive at least for
the tensile $1$-branes (strings). The appropriate ansatz is
\ba\nl x^{q,s}(\tau,\s^1) &=&
C^{q,s}(z^1\pm2\lambda^0 T_1\tau) + y^{q,s}\left(\tau\right),
\\ \nl x^M (\tau,\s^1)&=&y^M (\tau)\h\mbox{for}\h M\neq q,s.\ea
The corresponding solutions for $y^M$ are
\ba\nl y^q &=& y^q_0
\mp C \int_{y^r_0}^{y^r}du\frac{\left(C^q g^0_{qs}+C^s
g^0_{ss}\pm2\lambda^0T_1\frac{C^qh^0}{C}\right)}{h^0 U^{1/2}_0},
\\ \nl y^s &=& y^s_0
\pm C \int_{y^r_0}^{y^r}du \frac{\left(C^q g^0_{qq}+C^s
g^0_{qs}\mp2\lambda^0T_1\frac{C^sh^0}{C}\right)}{h^0 U^{1/2}_0},
\\ \nl y^{\alpha} &=& y^{\alpha}_0
\pm C^{\gamma}\int_{y^r_0}^{y^r}\frac{du}{U^{1/2}_0}
\exp\left(-\int dg^0_{\gamma\beta}g^{0\beta\alpha}\right),
\\ \nl y^k &=& constants,\h \tau = \tau_{0} \pm
\int_{y^r_0}^{y^r}\frac{du}{U^{1/2}_0} ,
\\ \nl U_0 &=& -\frac{D^0_k}{g_{rr}^0 g^0_{kk}} ,
\h D^0_{k} = g^0_{kk}
\left\{\left[\frac{C^2}{h^0}+\left(2\lambda^0T_1\right)^2\right]G^0
+ V^0 \right\}.\ea
It is evident that taking the limit $T_1\to 0$ in the above expressions,
we obtain our null string solution with $F(z^1)=z^1$.

Let us finally note that there exists another ansatz
which leads to the same type of exact solutions and it is
\ba\nl x^q (\tau,\sigma^a)&=&C^q F(z^a) + y^q (\sigma),\h
x^s (\tau,\sigma^a)=C^s F(z^a) + y^s (\sigma),\\ \nl x^M
(\tau,\s^a)&=&y^M (\sigma)\h\mbox{for}\h M\neq q,s,\ea
where $\sigma$ is one of the world-volume coordinates
$\sigma^1,\ldots,\sigma^p$. The corresponding tensile string ansatz
is obvious.

The author would like to thank J. Gamboa for the useful information.


\end{document}